\documentclass[%
 reprint,
 amsmath,amssymb,
 aps,pra
]{revtex4-2}

\usepackage{braket}
\usepackage{graphicx}% Include figure files
\usepackage{dcolumn}% Align table columns on decimal point
\usepackage{bm}% bold math

\begin{document}

\preprint{APS/123-QED}

\title{Dynamic rephasing in a telecom warm vapor quantum memory}

\author{I. Maillette de Buy Wenniger$^{1,\dagger}$, P. Burdekin$^{1,\dagger}$, S. Zhang$^1$, M. J. Rasiah$^1$, A. Rastogi$^1$, O. T. P. Schmidt$^{1,2}$, P. M. Ledingham$^{3}$, I. A. Walmsley$^1$, S. E. Thomas$^{1,4}$}
\email{sarah.thomas@eng.ox.ac.uk \\$^\dagger$These authors contributed equally to this work}
\affiliation{$^1$Department of Physics, Imperial College London, London, SW7 2BW, UK\\
$^2$Department of Physics, ETH Z\"urich, 8092 Z\"urich, Switzerland \\
$^3$Optoelectronics Research Centre, University of Southampton, Southampton SO17 1BJ, UK\\
$^4$Department of Engineering Science, University of Oxford, Oxford OX1 3PJ, UK }

\date{\today}

\begin{abstract}
 The Off-Resonant Cascaded Absorption (ORCA) protocol in warm atomic vapors offers a scalable platform for high-bandwidth, low noise quantum memories, but its coherence time is fundamentally limited by Doppler-induced dephasing. We introduce and experimentally demonstrate a dynamic rephasing protocol that counteracts Doppler dephasing in a telecom-band ORCA quantum memory. By transferring the stored excitation to an auxiliary shelving state, we effectively reverse the accumulated Doppler phase and extend the storage time by a factor of 50 while preserving the memory’s GHz bandwidth and low noise. Using this protocol, we then demonstrate on-demand storage and retrieval of four independent time-bin modes within a single warm vapor memory -- showing that Doppler dephasing can alternatively be harnessed for high-dimensional temporal mode processing. Our results establish rephasing in warm atomic vapors as a viable route toward high-bandwidth, temporally multiplexed quantum memories operating at room temperature.
\end{abstract}

\maketitle

\section{\label{sec:intro}Introduction}

The next-generation of quantum technologies will harness distributed quantum entanglement to link quantum processing nodes on both global and local scales. In these architectures, light forms a natural carrier of quantum information. However, losses in optical fiber and free-space transmission make scaling a formidable challenge. Photonic quantum memories -- devices that coherently store and retrieve quantum states of light on-demand -- play a key role by enabling the storage and manipulation of entangled photons. These capabilities allow quantum operations to be synchronized and distributed across local photonic quantum processors and quantum networks, while also serving as essential building blocks for long-distance quantum repeaters~\cite{Slussarenko2019, Gisin2007, Heshami2016}.

Among the various platforms, telecommunication-compatible quantum memories~\cite{Thomas2023, ThomasScience2024, Jin2015, Jiang2023, Askarani2019} are particularly interesting since they can be directly integrated within existing optical-fiber infrastructure operating at wavelengths where transmission losses are minimized. An alternative approach is to interface quantum memories operating outside the telecommunication-band with quantum frequency conversion (QFC)~\cite{Arensktter2023}. However, QFC introduces additional optical interfaces, each with finite efficiency and added noise, whose effects compound over large-scale quantum networks. In contrast, telecom-compatible memories remove this overhead entirely, providing a more direct and potentially scalable route toward long-distance quantum communication.\\

Warm atomic vapor memories provide a practical route toward such systems, combining relatively simple implementation with broadband operation. In particular, the Off-Resonant Cascaded Absorption (ORCA) protocol enables efficient, low-noise storage at telecommunication wavelengths, whereby the photonic quantum state is mapped to a collective atomic coherence between the ground and doubly excited states~\cite{Thomas2023, Kaczmarek2018}. However, the storage time is fundamentally limited by Doppler-induced dephasing: atoms with different velocities accumulate phase at different rates, leading to rapid decay of the collective coherence and strongly suppressing retrieval efficiency. In previously reported telecom-ORCA memories~\cite{Thomas2023, ThomasScience2024} this restricted the storage time to approximately one nanosecond -- far too short for quantum networking applications and insufficient to support temporal multiplexing.

At the same time, rapid dephasing causes photonic states separated by more than the dephasing time to evolve into orthogonal collective modes when stored in the memory. Reversing this dephasing would therefore enable temporally multimode operation. The ability to control Doppler dephasing while preserving the high bandwidth and low-noise performance of ORCA is thus key for enabling scalable, temporally multimode quantum networks.\\ 

Several approaches have been developed to mitigate Doppler-induced dephasing in atomic ensembles~\cite{Finkelstein2021, Li2025, Jiao2025}. One strategy is to suppress dephasing by continuously dressing the collective coherence with off-resonant fields, thereby engineering a protected state whose phase evolution is insensitive to atomic motion~\cite{Finkelstein2021}. While such approaches can effectively prevent dephasing, their application to the telecom-ORCA system would require strong optical fields with large detunings, which is challenging to implement. Furthermore, suppressing dephasing would remove the ability to use Doppler dephasing to realize temporally multimode operation.

\begin{figure*}[t]
\includegraphics[width=\linewidth]{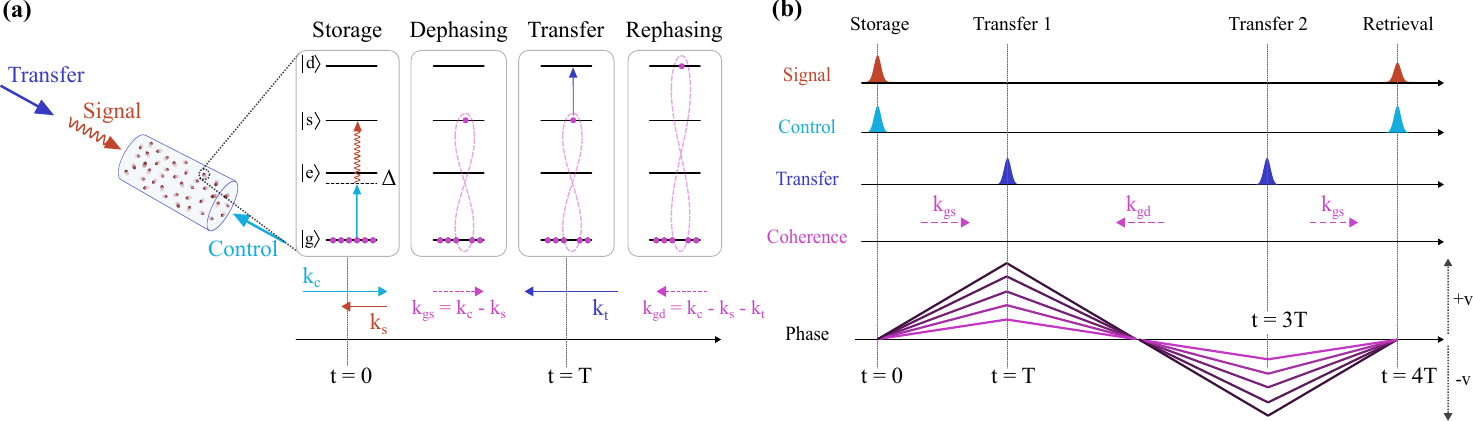}
\caption{Dynamic rephasing in an ORCA quantum memory. (a) At $t=0$, a weak optical field (red) is stored in an atomic ensemble via a two-photon process mediated by a strong control field (light blue), creating a stationary coherence between ground and doubly excited states, where each velocity class $v$ accumulates phase at the rate $k_{gs}~v$. At $t=T$, a transfer field (dark blue) maps the stored excitation to shelving state $\ket{d}$, where each velocity class in the ensemble accumulates phase at the rate $-k_{gd}~v$. (b) Pulse sequence diagram and atomic phase evolution for a single velocity class, where a shelving state $\ket{d}$ and transfer field direction have been chosen such that the transfer field reverses the direction of phase accumulation, i.e. $k_{gs} = -k_{gd}$. A second transfer field at $t=3T$ enables efficient retrieval at time $t=4T$ when the atoms are in phase again.  \label{fig:1}}
\end{figure*}

Instead, rather than continuously suppressing Doppler dephasing, here we present a protocol that reverses Doppler-induced dephasing in a warm rubidium (Rb) vapor ORCA memory operating in the telecom-band. By coherently transferring the stored excitation to an additional shelving state, we reverse the Doppler dephasing and extend the memory lifetime by more than an order of magnitude while preserving the room-temperature operation and GHz bandwidth. Moreover, we show that this approach enables the independent storage of multiple time bins, which can be rephased, manipulated and retrieved on-demand. In this way, we demonstrate that Doppler dephasing, previously a fundamental limitation to the memory lifetime, can be harnessed to realize temporal multimode operation, providing a route towards high-rate quantum communication and multiplexed repeaters.

\section{Telecom ORCA and Doppler dephasing}

The memory operation is based on the ORCA~\cite{Kaczmarek2018, Thomas2023} protocol with the addition of a novel rephasing step, see Fig.~\ref{fig:1}(a). The ORCA quantum memory comprises two counter-propagating optical fields -- a weak signal field and strong control field -- that are in two-photon resonance with an optical ladder transition of an atomic ensemble while being frequency detuned from the intermediate transition by $\Delta$. 

The state of the atomic ensemble after absorption of the signal can be written as: 
\begin{equation}
    \ket{\psi}_\mathrm{gs} = \frac{1}{\sqrt{N}} \sum_{j=1}^N e^{i\vec{k}_\mathrm{gs}\cdot\vec{v}_jt}\ket{g_1,\dots,s_j,\dots,g_N},
\end{equation}
\noindent where $\ket{g(s)_j}$ labels the ground (storage) state of the $j^\mathrm{th}$ atom, $\vec{v}_j$ is its velocity, $N$ is the total number of atoms, and $\vec{k}_\mathrm{gs}=\vec{k}_c - \vec{k}_s$ is the wavevector mismatch of the counter-propagating control and signal fields. Upon application of a second control pulse after some storage time, the collective atomic coherence is retrieved as an optical field at the signal wavelength, completing the memory operation.

ORCA quantum memories are typically implemented in warm atomic ensembles to ensure a large $N$, and therefore a large collective enhancement of the light-matter interaction. However, in warm atomic ensembles, the distribution of atomic velocities in combination with the wavevector mismatch, results in velocity-dependent phase accumulation across the ensemble, see lower panel Fig.~\ref{fig:1}(b). This causes the collective stored excitation in the memory (the orbital wave), $\ket{\psi}_\mathrm{gs}$, to dephase rapidly, reducing the retrieval efficiency and limiting the maximum storage time of the memory.

\section{Dynamic rephasing}
To rephase the atomic coherence, and thereby extend the maximum storage time, we apply an additional field (transfer field) at time $t=T$, between the storage and retrieval process. This transfer field resonantly couples to the atomic transition $\ket{s} \rightarrow \ket{d}$, where  $\ket{d}$ is an additional higher lying energy level of the atoms (Fig.~\ref{fig:1}(a)). For a pulse area corresponding to a $\pi$-pulse, the population is fully transferred to $\ket{d}$, and the initial orbital wave $\ket{\psi}_\mathrm{gs}$ is mapped onto a coherence between $\ket{g}$ and $\ket{d}$. The propagation direction of the transfer field is chosen such that the net wavevector is 
\begin{equation}
    \vec{k}_\mathrm{gd} = \vec{k}_\mathrm{c} - \vec{k}_\mathrm{s} - \vec{k}_\mathrm{t}, 
\end{equation}
which can have a sign opposite to $\vec{k}_\mathrm{gs}$ depending on the magnitude and direction of $\vec{k}_\mathrm{t}$. Consequently, each atomic velocity class accumulates phase in the opposite direction and the collective coherence rephases at a later time. 

\begin{figure*}[t]
\includegraphics[width=1\textwidth]{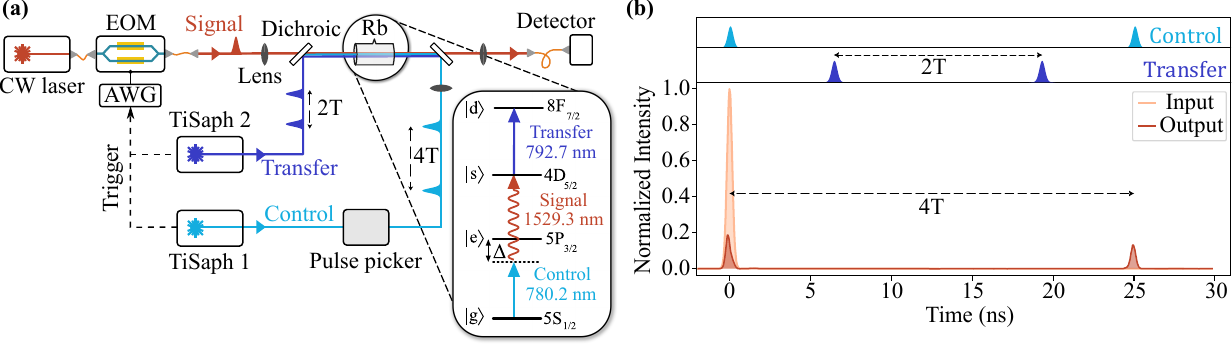}
\caption{(a) Experimental setup for the warm rubidium (Rb) vapor memory. Three synchronized lasers provide the pulsed signal (attenuated), control (strong), and transfer fields. Inset: atomic energy level scheme for the dynamic rephasing telecom-ORCA memory protocol. (b) Experimental data showing storage and retrieval using the rephasing protocol. Shown in the lower panel, storage and retrieval of a weak telecom signal field using the control and transfer field timings, illustrated in the upper two panels. }\label{fig:2}
\end{figure*}

This principle is illustrated in Fig.~\ref{fig:1} which shows the phase evolution of different atoms (with different velocities, shaded lines) as a function of time, together with the timings of the control, signal and transfer fields. If the transition $\ket{s} \rightarrow \ket{d}$ is chosen such that $\vec{k}_\mathrm{gd} = - \vec{k}_\mathrm{gs}$, a transfer field at $t=T$ yields complete rephasing at $t=2T$. However, retrieval into the signal mode requires the coherence to first return to $\ket{\psi}_\mathrm{gs}$. This could be achieved with the simultaneous application of the transfer and control fields, though the combined AC Stark shifts would need to be accounted for. Alternatively, we let phase accumulate in state $\ket{d}$ for longer and apply a second transfer field at $t=3T$, which reverses the phase accumulation once more; after an additional time $T$, the atomic coherence is rephased in state $\ket{s}$ and applying a second strong control field maps the stored excitation back onto the optical signal field. We note that this protocol does not strictly require $\vec{k}_\mathrm{gd} = - \vec{k}_\mathrm{gs}$, only that the vector signs are opposite. If the magnitudes differ, the phase accumulation rates for $\ket{\psi}_\mathrm{gs}$ and $\ket{\psi}_\mathrm{gd}$ are different and pulse timings must be adjusted accordingly.\\

\section{Experiment \label{sec:Exp}}

We implement the rephasing protocol with the setup shown in Fig.~\ref{fig:2}(a), using a $7.5~$cm cell containing $96.9\%$ $^{87}$Rb atoms heated to $120^\circ$C. The counter-propagating control and signal fields are in two-photon resonance with the $5\mathrm{S}_{1/2}\rightarrow 4\mathrm{D}_{5/2}$ ($\ket{g}\rightarrow\ket{s}$) transition, while being detuned from the intermediate $5\mathrm{P}_{3/2}$ state ($\ket{e}$) by $\Delta = 6$~GHz. This corresponds to wavelengths of 1529.3~nm and 780.2~nm for the signal and control fields, respectively. 

The control pulses are derived from a mode-locked Ti:Sapphire laser (Spectra Physics Tsunami) operating at a repetition rate of $80~$MHz, with a FWHM of $330~$ps (bandwidth $1~$GHz). The signal field, also with a duration of $330~$ps, is carved from a continuous-wave telecom laser (Santec) using a fiber intensity modulator (iXBlue), driven by an arbitrary waveform generator (Tektronix) and RF amplifier (iXBlue). The signal pulses are synchronized to the control laser repetition rate and repeated at $10~$MHz.

Neutral density filters are used to attenuate the signal input to a mean photon number of $0.0040(9)$ per pulse. High-transmission-band and long-pass filters before detection provide 14 orders of magnitude suppression of background light at the control wavelength, resulting in an overall signal transmission of $23\%$ from the cell to the detectors. The signal is detected using superconducting nanowire single-photon detectors (Photon Spot) together with a time-to-digital converter (Swabian) to produce start-stop histograms.

To improve upon the previous memory lifetime of $1.10(2)$~ns~\cite{Thomas2023}, we introduce a transfer field co-propagating with the signal and resonant with the $4\mathrm{D}_{5/2}\rightarrow8\mathrm{F}_{7/2}$ transition ($\ket{s}\rightarrow\ket{d}$) at 792.7~nm. This yields $\vec{k}_\mathrm{gd} \approx - \vec{k}_\mathrm{gs}$, thereby effectively reversing the Doppler phase evolution. The transfer pulses are derived from another mode-locked Ti:Sapphire laser (Spectra Physics), with a FWHM of $330~$ps, repetition rate of $80~$MHz and synchronized with the control pulse laser. The beam waist diameters of the signal, control and transfer field at the center of the vapor cell are approximately $220~\mu$m, $250~\mu$m, and $300~\mu$m, respectively. We apply the pulse sequence shown in Fig.~\ref{fig:1}(b) with T = 6.25~ns, resulting in a total storage time of 25~ns.\\ 

The experimental results are presented in Fig.~\ref{fig:2}(b). The lower panel shows storage of the telecom signal (orange) and its retrieval after 25~ns. The efficiencies are obtained by integrating the detected counts within a $984$~ps temporal window, chosen to maximize the signal-to-noise ratio (SNR). For a control pulse energy of 12~nJ we measure a storage efficiency of $83.6(7)\%$. In the absence of the transfer pulses, no retrieved signal is observed at $T=25$~ns (i.e. when the readout control pulse is applied), and we measure a total memory efficiency of $0.009(4)\%$, confirming no retrieval from the memory due to Doppler-induced dephasing. In contrast, applying two transfer fields with pulse energies of 1.6~nJ (resulting in a $\pi$-pulse fidelity of $89.2(7)\%$, see Appendix~\ref{sec:Transfer Pulse Fidelity}) at $t=6.25$~ns and $18.75$~ns yields a total memory efficiency of $12.6(1)\%$. These results demonstrate that the rephasing protocol effectively counteracts Doppler-induced decoherence and extends the storage time by more than an order of magnitude compared to previous telecom-ORCA demonstrations~\cite{Thomas2023, ThomasScience2024}. The observed total efficiency is mainly limited by hyperfine beating, as discussed in the following section, combined with imperfect transfer pulse fidelity, non-optimized control pulses and finite excited state lifetimes ($84~$ns for $\ket{s} = 4\mathrm{D}_{5/2}$~\cite{Wang2014} and $370~$ns for $\ket{d} = 8\mathrm{F}_{7/2}$~\cite{Theodosiou1984}).\\

To evaluate the noise performance of the rephased ORCA memory, we measure the photon counts in the retrieval window with the signal field blocked. With only the control field present, we detect $0.2(1)\times 10^{-6}$ photons per pulse; when both control and transfer fields are applied, the noise level remains at $0.3(3)\times 10^{-6}$ photons per pulse. This confirms that the rephasing protocol introduces negligible additional noise. For an input mean photon number of $\mu_\mathrm{in}=0.0040(9)$, we measure a SNR of the retrieved light exceeding $10^{3}$. Since the SNR scales linearly with input photon number, this corresponds to an SNR exceeding $10^{5}$ at the single-photon level ($\mu_\mathrm{in}\sim1$), demonstrating that the rephased telecom ORCA operates in a regime suitable for single-photon storage and retrieval. Notably, this low-noise performance is in contrast to photon-echo-based memories in similar systems, which are typically limited by excess noise processes in warm vapors~\cite{Leung2024}.\\

\section{Hyperfine beating \label{sec:Hyperfine beating}}
Although an efficiency of $12.8(1)\%$ at 25~ns represents a significant improvement, it remains lower than expected. Based on the measured storage efficiency, transfer pulse fidelity, and finite excited-state lifetimes, we estimate an expected retrieval efficiency on the order of $\sim 20\%$. As we show below, this discrepancy can be understood by considering the full hyperfine and magnetic sublevel structure of $^{87}$Rb (see Appendix~\ref{sec:Hyperfine Beating Experiment} \& \ref{sec:Rephasing Simulations}). 

In our initial description, the atoms were treated as ideal four-level systems. In reality, the $\ket{s}$ and $\ket{d}$ states consist of hyperfine and magnetic sublevels. The hyperfine splittings are on the order of $\sim$10-100~MHz. However, Doppler broadening prevents spectrally resolving individual hyperfine states, and as a result, multiple excitation pathways with different transition strengths contribute simultaneously to the memory process. The stored atomic coherence therefore consists of a superposition of contributions associated with different transition frequencies, which accumulate phase at different rates. This leads to interference in time, manifesting as oscillations in the retrieval efficiency, commonly referred to as hyperfine beating.

\begin{figure*}[t]
\includegraphics[width=0.9\linewidth]{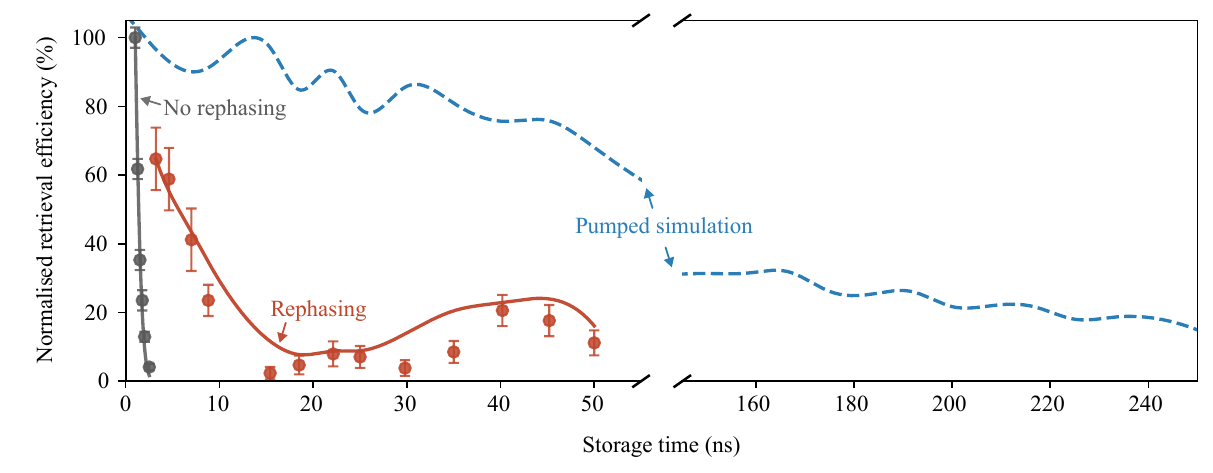}
\caption{The normalized retrieval efficiency as a function of storage time for three telecom-ORCA memory protocols. Black circles show the standard telecom-ORCA memory without rephasing, exhibiting rapid Gaussian decay with a characteristic lifetime of $800~\mathrm{ps}$. Red circles show the memory performance with the rephasing protocol applied, revealing oscillations due to hyperfine beating. The solid red curve shows results from numerical simulations including the full hyperfine and magnetic sublevel structure. The dashed blue curve shows the simulated retrieval efficiency for an optically pumped ensemble in the stretched state $\ket{F=2, m_{F}=+2}$, with all fields driving $\sigma^{+}$ transitions and chirped transfer pulses to improve fidelity. \label{fig:hyperfine_beating}}
\end{figure*}

We observe this interference experimentally when measuring the memory retrieval efficiency as a function of storage time (between 1-50~ns) with a modified setup (red circles in Fig.~\ref{fig:hyperfine_beating}; see Appendix~\ref{sec:Hyperfine Beating Experiment} for experimental details). For an average storage efficiency of $80(3)\%$, we observe pronounced oscillations in the retrieval efficiency with minima near $15$~ns and $30$~ns and maxima of varying amplitudes. These features are reproduced when performing simulations of the rephasing protocol with all hyperfine and magnetic sublevels, together with the full thermal velocity distribution (red solid line in Fig.~\ref{fig:hyperfine_beating}, see Appendix~\ref{sec:Rephasing Simulations}). For comparison, the black circles show the experimental results for the standard telecom-ORCA protocol without rephasing, as well as a Gaussian fit with characteristic lifetime $800~$ps. For all data and simulation the retrieval efficiency has been normalized to its value at the shortest storage time with no rephasing. 

To the best of our knowledge, no theoretical or experimental studies have reported the hyperfine splitting of the $8\mathrm{F}_{5/2}$ and $8\mathrm{F}_{7/2}$ transitions. In our simulations, we therefore set the magnetic dipole ($A$) and electric quadrupole ($B$) hyperfine constants as free parameters (see Appendix~\ref{sec:Rephasing Simulations}). We find good agreement with the experimental data for hyperfine splittings of $13(8)$ MHz, $15(4)$ MHz, and $14(4)$ MHz, corresponding to the $F=2\rightarrow3$, $3\rightarrow4$, and $4\rightarrow5$ intervals, respectively. These values should be regarded as effective parameters that reproduce the observed dynamics in our model, rather than precise measurements of the underlying hyperfine splitting. Minor deviations between the experimental data and simulations are attributed to imperfect control of optical field polarizations and transfer pulse energies. These results demonstrate that while Doppler dephasing can be effectively reversed using the rephasing protocol, hyperfine beating becomes the dominant limitation at longer storage times.

The impact of hyperfine beating can be mitigated by reducing the number of participating excitation pathways. One approach is to optically pump the atoms into a maximally stretched Zeeman sublevel of the $5\mathrm{S}_{1/2}$ $F=2$ ground state (e.g., $m_F=+2$) prior to memory operation. Through choosing the polarizations of all optical fields to drive $\sigma^{+}$ transitions in the atoms, the stored excitation can be restricted to a single hyperfine state within each manifold, effectively reducing the system to a four-level scheme. The resulting simulated retrieval efficiency is shown by the dashed blue line in Fig.~\ref{fig:hyperfine_beating}, using the same hyperfine splitting as above but assuming complete population preparation in $\ket{F=2,m_F=+2}$ and all fields polarized to $\sigma^{+}$. In this case, the decay in retrieval efficiency is consistent with an exponential decay with a $1/e$ lifetime of $140$~ns, limited solely by the radiative lifetimes of the $4\mathrm{D}_{5/2}$ and $8\mathrm{F}_{7/2}$. The observed residual oscillations are attributed to imperfect transfer pulses and rephasing.

An alternative scheme is to apply a magnetic field to lift the degeneracy of the Zeeman sublevels, allowing spectral selectivity of individual hyperfine states~\cite{Srivathsan2025}.  Further improvements could be achieved by transferring the stored coherence to a long-lived hyperfine ground state. Ground-state warm vapor memories have shown lifetimes of several milliseconds, limited by spin-exchange collisions between the atoms. Further extension to hundreds of milliseconds has been demonstrated by using a scheme insensitive to spin-exchange collisions~\cite{Katz2018}, though the randomization of atom velocities would prevent rephasing after long storage times. Instead the atomic coherence would need to be rephased before mapping to the ground state. Nevertheless, ground state mapping is a promising direction for extending our approach toward long-lived room-temperature telecom quantum memories.

\begin{figure}[t]
\includegraphics[width=\linewidth]{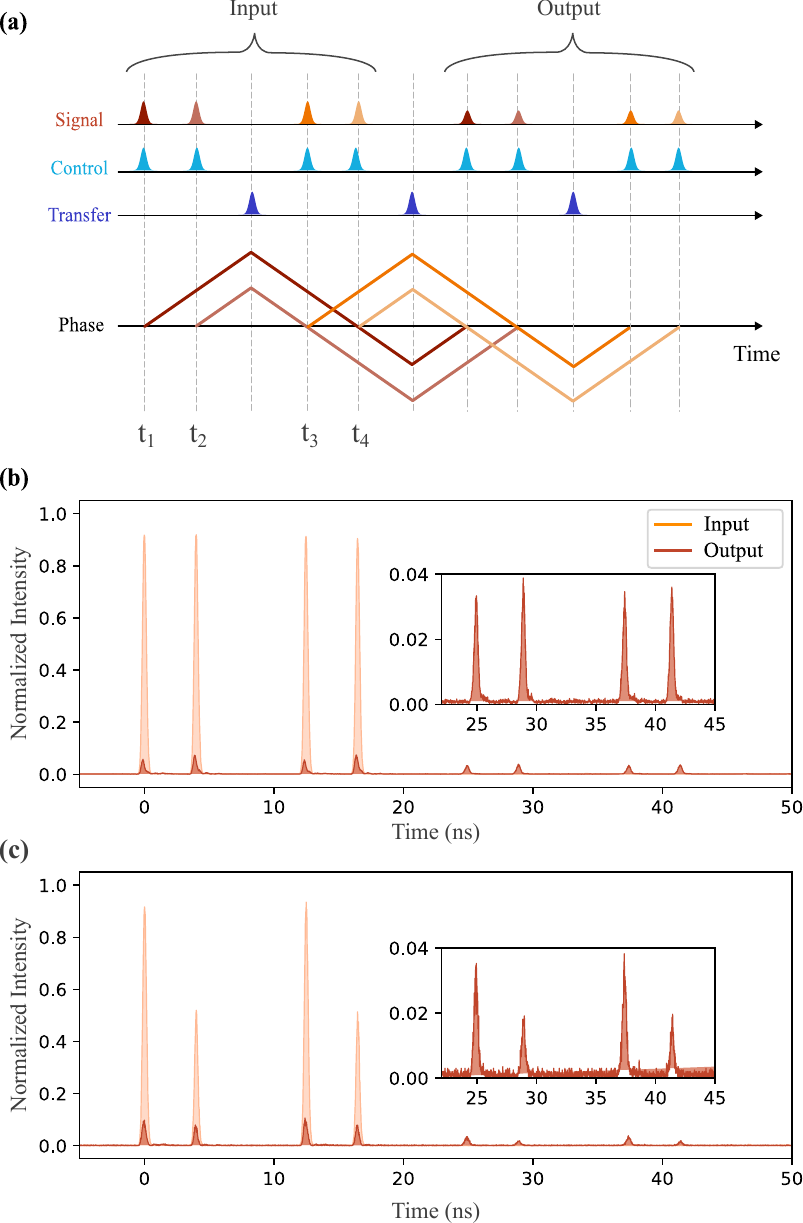}
\caption{Multimode storage in the time-bin basis. (a) Pulse sequence used to store four temporally separated input signals and retrieve them after a total storage time of 25~ns. The corresponding phase evolution of a single velocity class for each collective coherence during the rephasing protocol is indicated schematically. (b) Experimental data demonstrating storage and retrieval of four independent time-bin modes with equal input amplitudes. (c) Storage and retrieval of four time-bin modes with unequal input amplitudes, demonstrating preservation of relative mode weights.\label{fig:multimode}}
\end{figure}

\section{Multimode capacity}

This rephasing protocol not only extends the storage time of the telecom-ORCA quantum memory, but also enables multimode storage in the time-bin basis. Crucially, once a stored coherence has dephased it can no longer be retrieved by a control field without active rephasing. Instead, a subsequent control field can be used to store a signal field at a later time as an independent collective coherence. 

To illustrate this principle, consider two signal pulses arriving in time bins $t_1$ and $t_2$, separated by a delay greater than the Doppler dephasing time $t_\mathrm{deph} = 1.1$~ns, see Fig.~\ref{fig:multimode}(a). A control pulse at $t_1$ stores the first time bin as a collective coherence $\ket{\psi}_{gs}^{(1)}$. After a time exceeding $t_\mathrm{deph}$, this coherence has dephased and cannot be retrieved by a later control pulse. A control pulse arriving at time $t_2$ therefore stores the second input signal (at $t_2$) as an independent coherence $\ket{\psi}_{gs}^{(2)}$. The resulting two coherences can be rephased independently, and thus retrieved as two separate signals. This argument generalizes to an arbitrary number of temporal modes, provided that the time between subsequent time bins is greater than the memory dephasing time. Transfer pulses can either be applied after a sequence of input signals to collectively map all stored coherences to the shelving state, or in between input time bins. In this way, Doppler dephasing becomes a resource for multimode storage in warm vapor memories. \\

We experimentally demonstrate multimode storage and retrieval of four independent time bins ($t_1, t_2, t_3, t_4$), as shown in Fig.~\ref{fig:multimode}. Two signal pulses at $t_1=0$~ns and $t_2=4$~ns are stored as collective coherences $\ket{\psi}_\mathrm{gs}^{(1)}$ and $\ket{\psi}_\mathrm{gs}^{(2)}$, respectively. Since their separation exceeds the Doppler dephasing time, the first excitation is not retrieved at $t_2$. A transfer pulse at $6.25$~ns maps both coherences to the shelving state, creating $\ket{\psi}_\mathrm{gd}^{(1)}$ and $\ket{\psi}_\mathrm{gd}^{(2)}$.
Two further input signals at $t_3 = 12.5$~ns and $t_4 = 16.5$~ns generate coherences $\ket{\psi}_\mathrm{gs}^{(3)}$ and $\ket{\psi}_\mathrm{gs}^{(4)}$. At this stage, four independent modes are stored in the ensemble. A second transfer pulse at $18.75$~ns swaps the populations between $\ket{s}$ and $\ket{d}$, thereby (again) reversing phase accumulation for all coherences. Subsequent control pulses at $25$~ns and $29$~ns retrieve the first two time bins, while a final transfer pulse followed by two additional control pulses enables retrieval of the remaining two signals. Each time bin is thus stored for $25$~ns and retrieved independently, with an approximately uniform internal efficiency of $4.0(2)\%$. Note that the efficiency is lower here than in Fig.~\ref{fig:2} due to lower energy per control pulse. In addition, we show in Fig.~\ref{fig:multimode}(c) that the memory preserves the relative weights of time bins -- an essential requirement for the storage of time bin qubits. To demonstrate this, we repeat the experiment, but vary the input pulse intensities. The integrated counts for each input pulse, normalized to the first, are $[1, 0.517(4), 1.002(7), 0.517(4)]$. Each time bin is retrieved again after $25$~ns, and we measure the relative integrated counts for each bin, normalized to the first time bin, giving $[1, 0.54(2), 1.07(4), 0.54(2)]$, demonstrating that the rephased ORCA memory preserves the amplitude of time bin-encoded qubits.\\

In the present experiment, the number of stored modes is limited to four due to technical constraints in the pulse generation. The control and transfer pulse sequences are produced by splitting pulses from mode-locked lasers using free-space Mach–Zehnder interferometers, resulting in a fixed number of temporally separated pulses. Extending this approach to a larger number of modes would require additional interferometric stages, leading to increased optical complexity and a rapid reduction in pulse energy (by at least a factor of four per splitting stage), which limits the achievable control and transfer pulse energies and therefore attainable memory efficiency. These limitations can be overcome by generating the laser pulses through electro-optic modulation together with optical amplification, which when combined with mitigation of hyperfine beating would enable significantly larger multimode capacities in future implementations.

Beyond these technical constraints, the temporal multimode capacity of the memory is fundamentally determined by the ratio of the minimum separation between adjacent time bins (set by the dephasing time) and the memory lifetime. In our system, negligible cross-talk between modes requires a separation of approximately $3~t_\mathrm{deph}$, while the predicted storage time in absence of hyperfine beating is $140$~ns, yielding an estimated capacity on the order of $50$ independent time bin modes. This could be combined with additional degrees of freedom such as spectral, temporal, or spatial modes, to generate a highly multimode quantum memory in a single warm vapor cell.\\

\section{Discussion}
Temporally multiplexed storage has previously been demonstrated in warm atomic ensembles. For example, gradient echo memories have shown storage of up to 20 temporal modes~\cite{Hosseini2011}. In such systems, the control field power must be reduced to prevent premature retrieval of stored excitations, which in turn limits the achievable memory efficiency as the number of modes increases. In contrast, in our system Doppler-induced dephasing intrinsically suppresses unwanted readout. This enables high-bandwidth operation while maintaining high efficiency, and thus provides a fundamentally different route to scalable temporal multiplexing in warm vapor memories.

In addition, the ability to store independent temporal modes using the rephased ORCA protocol naturally extends to controlled manipulation of time bins within the memory. By tailoring the timing of the transfer and readout control pulses (see Appendix~\ref{sec:Time-bin Processor}), the rephasing sequence can be engineered to retrieve modes in a programmable order, enabling temporal reordering and on-demand synchronization within a single device. Furthermore, adjusting the transfer pulse area to achieve partial population transfer between $\ket{s}$ and $\ket{d}$ -- for instance via a $\pi/2$ pulse -- realizes an effective beam-splitter interaction between stored temporal modes, allowing for interference and phase-sensitive operations directly in the memory. In this way, the rephased ORCA protocol provides not only multimode storage, but also a flexible platform for in-memory temporal processing of photonic quantum states.\\

More generally, the dynamic rephasing protocol demonstrated here can be viewed within the broader context of techniques for reversing inhomogeneous dephasing. It bears conceptual similarities to spin-echo protocols~\cite{Hahn1950}, where a $\pi$-pulse inverts the phase evolution, as well as to photon-echo-based memories such as gradient echo memory (GEM)~\cite{Sparkes2013, Hosseini2011, Glorieux2012, Lauritzen2011, Leung2024}, in which a controllable inhomogeneous broadening is reversed to achieve rephasing. However, these approaches typically operate in narrowband regimes (MHz or below), and, in the case of solid-state implementations at telecom wavelengths~\cite{Lauritzen2011}, require cryogenic conditions. In contrast, our protocol operates in a warm atomic vapor and supports GHz-bandwidth signals. 

Finally, we also note that our protocol differs fundamentally from schemes based on continuous dressing~\cite{Finkelstein2021}, in which Doppler dephasing is suppressed through through strong off-resonant AC Stark shifts. Our approach instead dynamically reverses the accumulated phase without requiring continuous fields, and allows for the storage of multiple independent time-bin modes. Moreover, unlike related $\pi$-pulse rephasing schemes demonstrated in cold Rydberg ensembles~\cite{Jiao2025}, our protocol relies on inversion of the collective wavevector and is therefore insensitive to the optical phase of the driving fields. Together, these features establish dynamic rephasing in ORCA as a distinct and complementary approach for achieving long-lived, active broadband multimode quantum memories.

\section{Conclusion}

We have introduced and experimentally demonstrated a dynamic rephasing protocol that counteracts Doppler-induced dephasing in a warm vapor ORCA quantum memory. By reversing the accumulated phase evolution through coherent population transfer to an additional shelving state, we extend the storage time of the telecom memory by a factor of 50 while preserving its intrinsically high bandwidth and low-noise performance. Beyond extending the memory lifetime, the rephasing scheme enables new operational capabilities, including temporal multimode storage, which we demonstrate through the storage and retrieval of four independent time-bin modes. In addition, the protocol provides a route toward more advanced temporal processing, such as programmable reordering and interference of time-bin signals. These capabilities position the rephased telecom ORCA memory as a versatile processor for high-bandwidth time-bin modes, combining storage, re-ordering and interference in a single device.

\vspace{3mm}

\textbf{Acknowledgments} -- This work was supported by Innovate UK (Grant No. 10102696) and the European Union's Horizon 2020 Research and Innovation Programme Grant No. 731473 Quiche. This work was also funded within the QuantERA II Programme which has received funding from the European Union's Horizon 2020 research and innovation programme under Grant No. 101017733 (EQSOTIC). 

We acknowledge support from the EPSRC Integrated Quantum Networks Hub (IQN, Grant No. EP/Z533208/1). P.M.L. acknowledges support from UK Research and Innovation (Future Leaders Fellowship, Grant No. MR/V023845/1).

\bibliography{RephasedORCA}
\clearpage
\newpage
\onecolumngrid
\appendix

\setcounter{figure}{0} \renewcommand{\thefigure}{A.\arabic{figure}}
\setcounter{equation}{0} 
\renewcommand{\theequation}{A.\arabic{equation}}

\title{Appendix}
\begin{center}
    \large{Appendix}
\end{center}
\section{Transfer pulse fidelity \label{sec:Transfer Pulse Fidelity}}

\begin{figure}[h]
\includegraphics[width=0.5\linewidth]{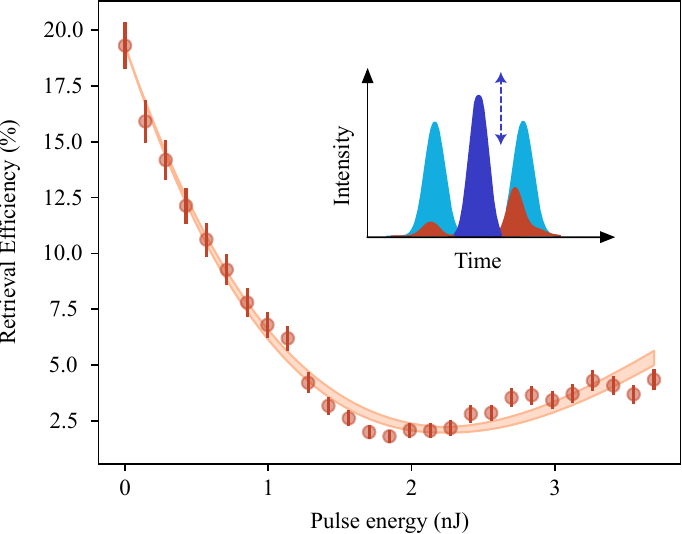}
\caption{Retrieval efficiency (red circles) as a function of transfer pulse energy for the pulse sequence shown in the inset. Inset: timing diagram of the signal (red), control (light blue), and transfer (dark blue) fields, with the dark blue arrow indicating that only the transfer pulse energy is varied. The orange band shows a $\cos^2$ fit to the data, with the shaded region indicating the uncertainty obtained from bootstrap resampling.\label{fig:TORCA_transfer_pi_pulse}}
\end{figure}

To determine the transfer pulse energy required to achieve population inversion from state $\ket{s}$ to $\ket{d}$, we use the setup described in Section~\ref{sec:Exp} and set the storage time to $1~$ns, resulting in a retrieval efficiency of $19.3(5)\%$. A transfer pulse is then applied between the write and read control fields (500~ps after storage), and its pulse energy is varied while monitoring the retrieval efficiency. An illustration of the pulse sequence is shown in the inset of Fig.~\ref{fig:TORCA_transfer_pi_pulse}. The signal, control, and transfer pulses all have a FWHM of $330~$ps, and the integration window used to compute the efficiencies was $1~$ns. Fig.~\ref{fig:TORCA_transfer_pi_pulse} shows the measured retrieval efficiency as a function of transfer pulse energy (red circles). The error bars include contributions from Poissonian counting statistics, as well as slow drifts in the memory coupling strength, estimated from fluctuations in the measured storage efficiency over the duration of the experiment.

The data are fit with a $\cos^2$ model (orange band) describing Rabi oscillations as a function of pulse area (proportional to square root of pulse energy). To estimate uncertainties in the fitted parameters, we perform a bootstrap resampling procedure, in which synthetic datasets are generated by sampling each data point from a normal distribution defined by its measured mean and standard deviation. The fitting procedure is repeated for each resampled dataset, yielding distributions for the fit parameters. From this analysis, we extract a $\pi$-pulse fidelity of $89.2(7)\%$ at a pulse energy of $2.24(3)~$nJ.

The deviation from unit fidelity is primarily attributed to spatial inhomogeneity of the transfer field, leading to variations in the Rabi frequency across the atomic ensemble, as well as velocity-dependent detunings arising from Doppler broadening, which prevents complete population inversion. The fidelity could be improved by a more uniform beam profile and by employing frequency-chirped or adiabatic transfer pulses to mitigate sensitivity to detuning. We note that the optimal transfer pulse energy depends on experimental conditions such as relative beam alignment. As a result, the pulse energy used in the main text differs slightly from the value reported here, as it was optimized independently for the corresponding dataset. The extracted $\pi$-pulse fidelity remains representative of the achievable transfer efficiency under typical operating conditions.

\section{Hyperfine beating experiment \label{sec:Hyperfine Beating Experiment}}

We characterize the hyperfine beating by measuring the retrieval efficiency as a function of storage time (see Fig.~\ref{fig:hyperfine_beating}). Here, we use a slightly modified experimental setup (Fig.~\ref{fig:hyperfine_supp}(a)) compared to the setup in the main text. In this configuration, a longer vapor cell (14~cm) is used, with an ensemble temperature of approximately $70^{\circ}$C, and the beam waists are adjusted to approximately $200~\mu$m (signal), $380~\mu$m (control), and $350~\mu$m (transfer). The signal field is operated at $780.2~$nm and the control field at $1529.3~$nm. Exchanging the roles of the signal and control fields between the two transitions does not affect the hyperfine beating dynamics explored here, as the resulting atomic coherences involve the same manifolds. Further discussion of the differences between these configurations is given at the end of Appendix~\ref{sec:Rephasing Simulations}.

The control field is derived from a continuous-wave telecom laser (NKT Photonics), carved using a fiber intensity modulator (iXBlue) and amplified using an Erbium-Doped Fibre Amplifier (Thorlabs). This system enables precise electronic control over the timing of the readout pulse, whose arrival time is varied using a programmable arbitrary waveform generator (Tektronix). The signal field is similarly generated from a continuous-wave $780~$nm laser (Toptica) using a fiber intensity modulator (Optilabs), driven by an electrical pulse generator (Alnair Labs).The resulting signal pulses have a slightly longer full width at half maximum (FWHM) than in the main experiment, $500~\mathrm{ps}$ compared to $330~\mathrm{ps}$. Accordingly, the temporal integration window used to extract the detected counts is increased to $1.2~\mathrm{ns}$ in order to maximize the signal-to-noise ratio.

The transfer pulses are generated using two Ti:Sapphire pulsed lasers, each producing $330~$ps pulses at a central wavelength of $793$~nm. The signal, control, and transfer fields are all synchronized to a common $80~$MHz clock to ensure stable relative timing. The transfer pulses are combined on a polarizing beam splitter (PBS), followed by a half-wave plate (HWP) and a second PBS to ensure identical polarization and equal pulse energy in both arms. The relative timing between the two transfer pulses is controlled by introducing a motorized optical delay stage in one of the beam paths. This allows the separation between transfer pulses to be varied between $1.6$~ns and $12.5$~ns, corresponding to storage times $3.2$~ns to $25$~ns. To extend the achievable rephasing interval further, an additional Pockels cell is placed in one of the transfer beam paths, allowing selective suppression of pulses and enabling longer delays between successive transfer pulses. We measure the transfer pulse fidelity, using the same pulse sequence as shown in the inset of Fig.~\ref{fig:TORCA_transfer_pi_pulse}. We obtain a $\pi-$pulse fidelity of $83(1)\%$ for a transfer pulse energy of $2.10(3)~$nJ.

\section{Simulations of the rephasing protocol \label{sec:Rephasing Simulations}}

\subsection{Four-level system}

To model the rephasing scheme, we numerically solved the Maxwell–Bloch equations in Python. Since the rephasing experiment was performed with the signal field tuned to the $\ket{g}\rightarrow\ket{e}$ transition (at $780~$nm), we present the system of equations for this configuration. At the end of this section, we comment on the differences with the telecom-ORCA configuration, i.e. where the signal instead is tuned to the $\ket{e}\rightarrow\ket{s}$ transition. \\

Time evolution is computed using a fourth-order Runge–Kutta method, while spatial propagation is treated using a Chebyshev spectral method. For the four-level system used in the rephased ORCA, working in the reference frame co-moving with the signal field with proper time $\tau = t - z/c$, we can adiabatically eliminate the macroscopic atomic coherences between $\ket{g}$ and $\ket{e}$, as well as between $\ket{e}$ and $\ket{s}$ (due to large single-photon detuning) and we obtain the following equations of motion:
\begin{align}
    &\begin{aligned}
     \partial_{z}\mathcal{E}^{(\tau, z)}_{s} = \sum_{v}\left(\frac{+i\sqrt{d^{(v)}}\,\Omega_{c} S^{(\tau, z, v)}_{gs} - d^{(v)\mathcal{E}^{(\tau, z)}_{s}}}{\gamma_{e} + i\Delta^{(v)}_{s}}\right)
     \label{eq:4ORCAv_E}
    \end{aligned}\\
     &\begin{aligned}
    \partial_{\tau} S^{(\tau, z, v)}_{gs} = -\left(\gamma_{s}+i\Delta^{(v)}_{\rm II} + \frac{|\Omega_{c}|^{2}}{\gamma_{e}+i\Delta^{(v)}_{s}}\right)S_{gs}^{(\tau, z, v)} - \frac{i\sqrt{d^{(v)}}\,\Omega_{c}^{*}\mathcal{E}^{(\tau, z)}_{s}}{\gamma_{e}+i\Delta^{(v)}_{s}} - i\Omega_{d} S^{(\tau, z, v)}_{gd}
    \label{eq:4ORCAv_S}
    \end{aligned}\\
    &\begin{aligned}
    \partial_{\tau} S^{(\tau, z, v)}_{gd} = -\left(\gamma_{d}+i\Delta^{(v)}_{\rm III}\right)S^{(\tau, z, v)}_{gd} - i\Omega_{d}^{*} S^{(\tau, z, v)}_{gs}\, ,
    \label{eq:4ORCAv_D}
    \end{aligned}
\end{align}
where $\mathcal{E}^{(\tau, z)}_{s}$ denotes the telecom signal electromagnetic field under the paraxial approximation. The quantities $S^{(\tau, z, v)}_{gs}$ and $S^{(\tau, z, v)}_{gd}$ represent the macroscopic atomic coherences between states $\ket{g}$ and $\ket{s}$, and $\ket{g}$ and $\ket{d}$, respectively, evaluated at each time and spatial point $(\tau, z)$. The rotating wave approximation has been applied to the atomic coherences.

In all equations $v$ labels an atomic velocity class -- following the one-dimensional Maxwell-Boltzmann distribution $f(v, T)$ at temperature $T$ -- giving rise to the velocity-dependent optical depth $d^{(v)}= d~f(v, T) dv$ and detunings:
\begin{align*}
    \Delta_{s}^{v}&=\Delta_s + k_{s} v, & \Delta_{c}^{v}&=\Delta_c + k_{c} v & \Delta_{d}^{v}&=\Delta_d + k_{d} v \, ,
\end{align*}
for the signal, control and transfer fields respectively, with corresponding wavevectors $k_s$, $k_c$ and $k_d$. The two-photon detuning is therefore:
\begin{equation*}
    \Delta_{II}^{(v)} = \Delta_{s}^{v} + \Delta_{c}^{v}.
\end{equation*}
 Similarly the three-photon detuning is:
 \begin{equation*}
    \Delta_{III}^{(v)} = \Delta_{s}^{v} + \Delta_{c}^{v} + \Delta_{d}^{v}.
\end{equation*}
 The complex Rabi frequencies of the control and transfer fields are given by $\Omega_{c} = \Omega_{c}(\tau, z)$ and $\Omega_{d} = \Omega_{d}(\tau, z)$ respectively. The coherences $S^{(\tau, z, v)}_{gs}$ and $S^{(\tau, z, v)}_{gd}$ decohere due to spontaneous decay from the excited states $\ket{s}$ and $\ket{d}$, with spontaneous decay rates $\Gamma_{s}$ and $\Gamma_{d}$. Accordingly we approximate the coherence decay rates as $\gamma_{s}\approx\Gamma_{s}/2$ and $\gamma_{d}\approx\Gamma_{d}/2$.

\subsection{Full hyperfine and Zeeman sublevels}

In order to model the observed hyperfine beating, it is necessary to include the hyperfine and Zeeman sublevel manifolds for each of the states $\ket{g}$, $\ket{e}$, $\ket{s}$, and $\ket{d}$. This leads to a large set of coupled equations describing the atomic coherences. To improve computational efficiency we express the dynamical variables in tensor form with the relevant indices written explicitly in superscript brackets.

We index the hyperfine ground states with $g$, with corresponding magnetic sublevels $m_{g}$ . For the numerical results presented in Fig.~\ref{fig:hyperfine_beating} only the ground state $F=2$ is considered. The hyperfine states comprising state $\ket{e}$ are indexed by $j$, where again each $j$ has a set of magnetic sublevels indexed by $m_{j}$. The hyperfine and magnetic sublevels of $\ket{s}$ are indexed by $q$ and $m_{q}$, and those of $\ket{d}$ by $b$ and $m_{b}$. For example, the coherence $S_{gs}^{(\tau, z, g, m_{g}, q, m_{q}, v)}$ has indices relating to the time and space coordinates, together with indices which refer to a combination of sublevels in the states $\ket{g}$ and $\ket{s}$, and a final index to denote the velocity class of the atom. The optical depth now depends explicitly on the transition under consideration, $d^{(g, m_{g}, j, m_{j}, v, Q)}$, where the strength of the transition between the state indexed by $(g, m_{g})$ and state $(j, m_{j})$ driven by a signal field with polarization $Q$, is determined by the appropriate Clebsch–Gordan coefficients. The population of the ground-state sublevels is incorporated via the factor $\rho^{(g, m_{g})}$. Similarly, the Rabi frequency of the control and transfer fields are written as $\Omega^{(j, m_{j}, q, m_{q}, Q)}_{c} C_{c}^{(\tau, z, Q)}$ and $\Omega^{(q, m_{q}, b, m_{b}, Q)}_{d} C_{d}^{(\tau, z, Q)}$, respectively. The $\Omega$ terms contain the Clebsch-Gordan coefficients for the corresponding transitions, while the $C$ terms describe the spatiotemporal envelope and polarization of the classical driving fields. The signal electromagnetic field $\mathcal{E}$ likewise carries a polarization index $Q$.

The resulting set of coupled equations are:
\begin{align}
    \begin{aligned}
    &\partial_{z} \mathcal{E}^{(\tau, z, Q)} =i\sum_{g, m_{g}, j, m_{j}, v}\left(\sqrt{d^{(g, m_{g}, j, m_{j}, v, Q)} \rho^{(g, m_{g})}}\frac{\sum_{q, m_{q}}\left(\sum_{Q}\left(\Omega^{(j, m_{j}, q, m_{q}, Q)}_{c} C^{(\tau, z, Q)}\right) S^{(\tau, z, g, m_{g}, q, m_{q}, v)}_{gs}\right)}{\gamma_{e}+i\Delta^{(g, j, v)}_{s}}\right)\\ &-\sum_{g, m_{g}, j, m_{j}, v}\left(\frac{\sum_{Q}\left(d^{(g, m_{g}, j, m_{j}, v, Q)} \rho^{(g, m_{g})}\mathcal{E}^{(\tau, z, Q)}\right)}{\gamma_{e}+i\Delta^{(g, j, v)}_{s}}\right)
    \label{eq:E_hyperfine}
    \end{aligned}\\
    \begin{aligned}
    &\partial_{\tau} S^{(\tau, z, g, m_{g}, q, m_{q}, v)}_{gs} = -\Bigg[\gamma_{s} + i\Delta_{\rm{II}}^{(g, q, v)}\\ &- \sum_{j, m_{j}}\left(\frac{\sum_{Q}\left(\Omega^{(j, m_{j}, q, m_{q}, Q)}_{c} \left(C_{c}^{(\tau, z, Q)}\right)^{*}\right)\sum_{q, m_{q}}\left(\sum_{Q}\left(\Omega^{(j, m_{j}, q, m_{q}, Q)}_{c} C_{c}^{(\tau, z, Q)}\right)S^{(\tau, z, g, m_{g}, q, m_{q}, v)}_{gs}\right) }{\gamma_{e}+i\Delta_{s}^{(g, j, v)}}\right)\Bigg] \\ &-i \sum_{j, m_{j}}\left(\sum_{Q}\left(\Omega^{(j, m_{j}, q, m_{q}, Q)}_{c} \left(C_{c}^{(\tau, z, Q)}\right)^{*}\right)\frac{\sum_{Q}\left(\sqrt{d^{(g, m_{g}, j, m_{j}, v, Q)} \rho^{(g, m_{g})}}\mathcal{E}^{(\tau, z, Q)}_{s}\right)}{\gamma_{e}+i\Delta_{s}^{(g, j, v)}}\right)\\ & -i \sum_{b, m_{b}}\left(\sum_{Q}\left(\Omega^{(q, m_{q}, b, m_{b}, Q)}_{d} C_{d}^{(\tau, z, Q)}\right)S_{gd}^{(\tau, z, g, m_{g}, b, m_{b}, v)}\right)
    \end{aligned}\\
    \begin{aligned}
    &\partial_{\tau} S^{(\tau, z, g, m_{g}, b, m_{b}, v)}_{gd} = -\left(\gamma_{d}+i\Delta^{(g, b, v)}_{\rm III}\right)S^{(\tau, z, g, m_{g}, b, m_{b}, v)}_{gd} -i\sum_{q, m_{q}}\left(\sum_{Q}\left(\Omega^{(q, m_{q}, b, m_{b}, Q)}_{d} C_{d}^{(\tau, z, Q)}\right)^{*}S_{gs}^{(\tau, z, g, m_{g}, q, m_{q}, v)}\right)\, .
    \end{aligned}
\end{align}
The detunings now depend explicitly on the hyperfine levels involved, as well as on the atomic velocity class through the Doppler shift. For a given hyperfine level $F$, the energy relative to the unperturbed fine-structure transition is given by~\cite{Steck2001},
\begin{equation}
E_{HFS} = \frac{K A_{HFS}}{2} + \frac{B_{HFS}}{2} \left(\frac{3K(K + 1) - 4I(I + 1)J(J + 1)}{2I(2I - 1)2J(2J - 1)}\right),
\end{equation}
where
\begin{equation}
K = F(F + 1) - I(I + 1) - J(J + 1).
\end{equation}
Here, $A_{HFS}$ and $B_{HFS}$ denote the magnetic dipole and electric quadrupole hyperfine constants, respectively.

\begin{figure}[t]
\includegraphics[width=0.7\linewidth]{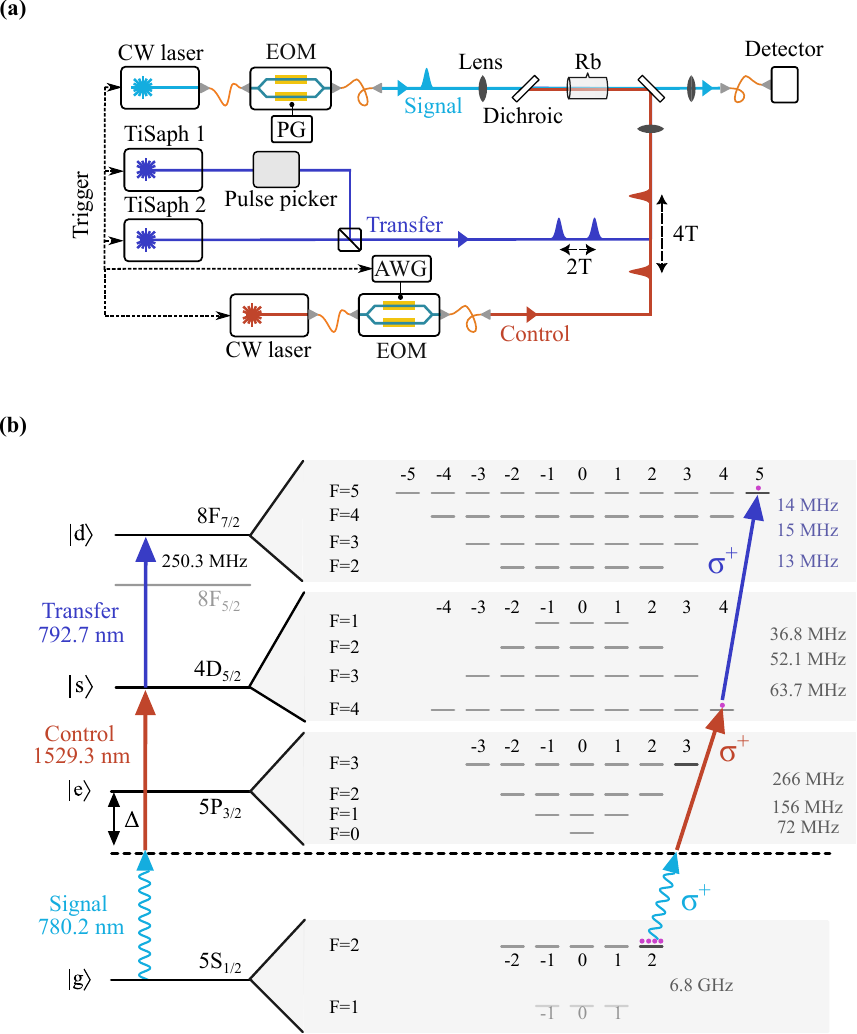}
\caption{(a) Modified experimental setup used to measure the retrieval efficiency as a function of storage time. The $780.2~\mathrm{nm}$ signal field is stored via a $1529.3~\mathrm{nm}$ control field, while the transfer field remains at $792.7~\mathrm{nm}$. (b) Atomic level configuration for the dynamic rephasing telecom-ORCA protocol. The left panel shows the relevant fine-structure states ($5\mathrm{S}{1/2}$, $5\mathrm{P}{3/2}$, $4\mathrm{D}{5/2}$, $8\mathrm{F}{7/2}$ and $8\mathrm{F}{5/2}$) and the optical couplings driven by the signal, control, and transfer fields. The right panel displays the full hyperfine and Zeeman sublevel structure included in the simulations. Here, the arrows indicate the reduced configuration obtained when all atoms are optically pumped into the stretched state $m_F=+2$ and all optical fields are polarized to drive $\sigma^{+}$ transitions, effectively restricting the dynamics to a single excitation pathway. The hyperfine splittings are indicated on the far right; for the $8\mathrm{F}{7/2}$ manifold, the splittings extracted from our numerical simulations are shown in blue.\label{fig:hyperfine_supp}}
\end{figure}

For $^{87}$Rb, the nuclear spin is $I = 3/2$. The initial storage state $\ket{s} = 4D_{5/2}$ has $J = 5/2$ and hyperfine levels $F = 1, 2, 3, 4$. The experimentally measured values of the hyperfine constants for this state are $A_{HFS} = -16.801(5)$~MHz and $B_{HFS} = 3.645(3)$~MHz~\cite{Wang2014}, where the negative sign of $A_{HFS}$ implies that the $F=1$ level lies highest in energy. The corresponding hyperfine structure is illustrated in Fig.~\ref{fig:hyperfine_supp}.

For the additional state used to rephase the stored coherence, $\ket{d} = 8F_{7/2}$, we have $J = 7/2$ and hyperfine levels $F = 2, 3, 4, 5$. To the best of our knowledge, no experimental measurements of the magnetic dipole and electric quadrupole hyperfine constants have been reported in the literature for this state. To determine suitable values of $A_{HFS}$ and $B_{HFS}$ for $\ket{d}$, we performed numerical simulations over a range of trial parameters using the Imperial High Performance Computing cluster CX3~\cite{ImperialRCS}, for the experimentally measured storage times shown in Fig.~\ref{fig:hyperfine_beating}. We do not have an accurate measure of the polarizations of the three fields when interacting with the atoms. As such, We choose the polarizations of the signal, control and transfer fields to be horizontal, vertical and vertical, respectively. Varying the polarizations of the fields would result in slightly different estimates for the hyperfine splitting.

To estimate $A_{HFS}$ and $B_{HFS}$, we computed the sum of squared residuals, $S$, between the simulated and experimental data over all storage times. To account for experimental uncertainties, each data point was sampled from a normal distribution defined by its measured mean and standard deviation. The metric $S$ was evaluated over the full parameter grid of simulated $A_{HFS}$ and $B_{HFS}$ values and subsequently interpolated to determine the location of the minimum. From this procedure, we obtained $A_{HFS} = 3.4(6)~$MHz and $B_{HFS} = -4(5)~$MHz. These values were then used to simulate the dynamics at finer temporal resolution, and the results are plotted in Fig.~\ref{fig:hyperfine_beating}. The corresponding estimated hyperfine splittings are shown in Fig.~\ref{fig:hyperfine_supp}. We do not intend for these values to be taken as the true hyperfine constants for the $8F_{7/2}$ state, but rather we include them here to support our conclusion of the beating shown in Fig.~\ref{fig:hyperfine_beating} originating from the splitting of the hyperfine levels.

In Fig.~\ref{fig:hyperfine_beating}, we additionally show the retrieval efficiency as a function of storage time, assuming that the atomic ensemble is initially prepared in the stretched spin state $m_F = +2$. This curve was obtained using the same simulation parameters as in the hyperfine beating analysis, but with the initial population distribution restricted entirely to the $m_F = +2$ Zeeman sublevel and all optical fields set to drive $\sigma^{+}$ transitions. In addition, we introduced a chirp to the transfer fields to improve the fidelity of each $\pi$-pulse to $99.1\%$ in simulation.

\subsection{Comment on the effect of swapping the control and signal field}

Swapping the roles of the signal and control fields does not alter the atomic states involved in the stored coherences, and therefore we do not expect the hyperfine beating dynamics to be affected. However, there are subtle differences in the write and read dynamics between the two configurations.

In particular, when the strong control field couples to the initially populated ground-state transition, even far detuned, it may induce additional atomic coherences. A further difference arises from whether the signal field couples to a populated transition. For a $780~\mathrm{nm}$ signal, the equations describing the electric field propagation (Eq.~\ref{eq:4ORCAv_E} and Eq.~\ref{eq:E_hyperfine}) contain a term proportional to the optical depth, in addition to the $\sqrt{d}$ coupling. The term proportional to the optical depth accounts for absorption and dispersion due to interaction with the populated transition. However, for the far-detuned signals employed here, this contribution is minimal. For a signal at $1529~\mathrm{nm}$, this term is absent.

\section{A dynamically rephased memory as a time-bin processor \label{sec:Time-bin Processor}}

\begin{figure}[t]
\includegraphics[width=0.7\linewidth]{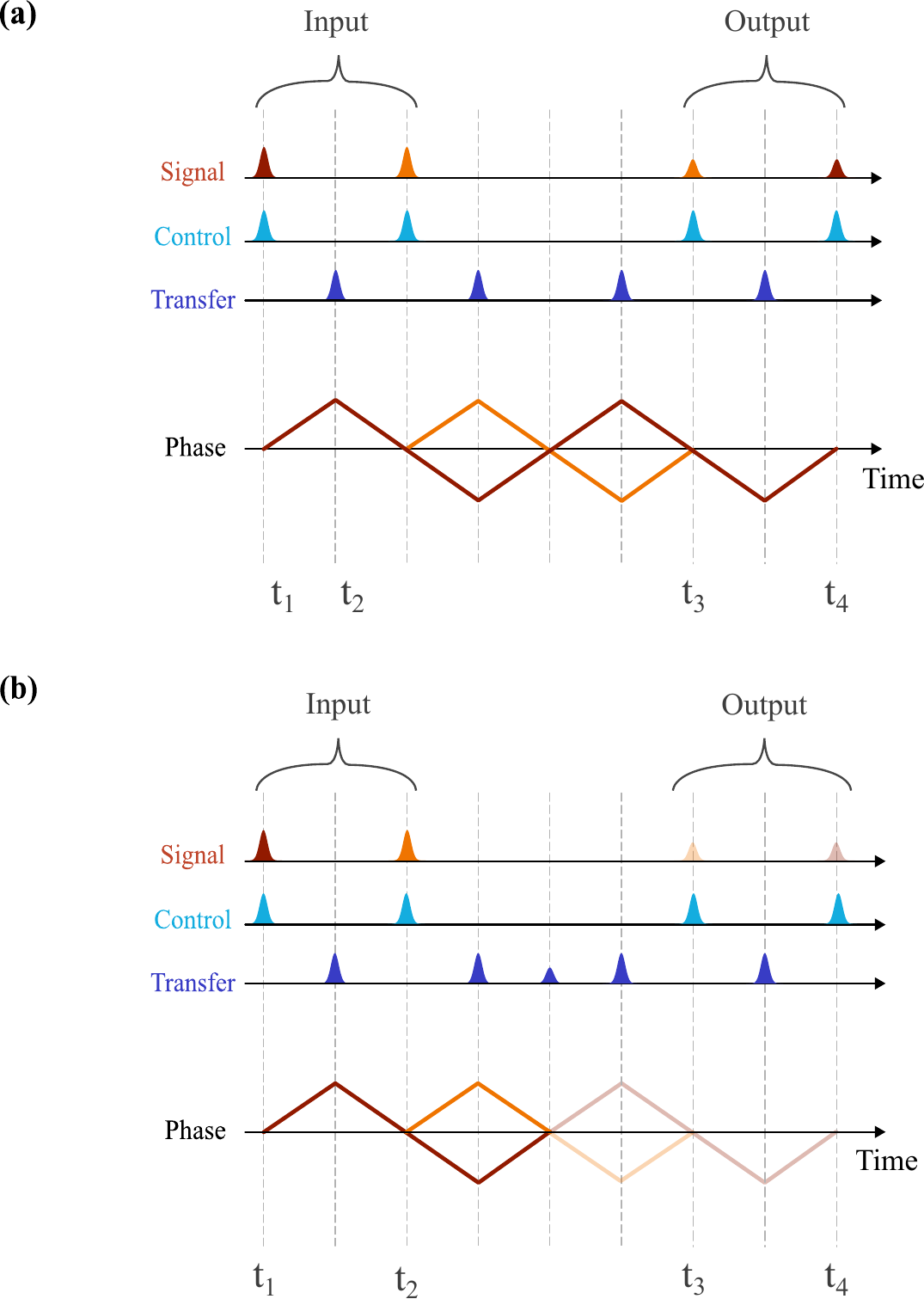}
\caption{Examples of pulse sequences enabling temporal-mode manipulation using the rephasing protocol, togehter with the phase evolution for a representative velocity class for each of the collective atomic coherences. (a) Pulse sequence for reversing the retrieval order of two stored time bins. Two inputs are stored at times $t_1$ and $t_2$, and a sequence of transfer and control pulses selectively rephases and retrieves the modes in reversed order. (b) Pulse sequence for interfering two stored time bins. A transfer $\pi/2$-pulse implements an effective beamsplitter between the two atomic coherences, analogous to a Mach--Zehnder interferometer in the time domain.
\label{fig:temporal_mode_manipulation}}
\end{figure}

In Fig.~\ref{fig:temporal_mode_manipulation}, we show two examples of control and transfer pulse sequences that enable manipulation of temporal modes within the memory. First, Fig.~\ref{fig:temporal_mode_manipulation}(a) illustrates a pulse sequence that reverses the retrieval order of two stored time bins. Two input pulses (red and orange) are stored at times $t_{1}$ and $t_{2}$, where $t_{2} - t_{1} \gg t_{\mathrm{deph}}$ with $t_{\mathrm{deph}}$ the Doppler dephasing time ($\sim 1~$ns in our system). Following storage, a transfer $\pi$-pulse is applied, mapping both collective coherences to the auxiliary state $\ket{d}$ and reversing their phase evolution. The phase evolution for a representative velocity class is shown for both coherences, with rephasing occurring when the phase crosses zero. A second transfer $\pi$-pulse maps the coherences back to the storage state $\ket{s}$, after which the two modes rephase at different times. A readout control pulse is applied at time $t_{3}$, when the coherence associated with the second input is rephased while the first remains dephased. As a result, only the second input is retrieved. Subsequently, an additional pair of transfer pulses rephases the coherence associated with the first input, which is retrieved at time $t_{4}$. In this way, the retrieval order of the stored time bins is reversed.

Second, the rephasing protocol enables interference between temporal modes within the memory. An example sequence is shown in Fig.~\ref{fig:temporal_mode_manipulation}(b). A first input pulse stored at time $t_{1}$ is transferred to $\ket{d}$ by a transfer $\pi$-pulse. A second input pulse is then stored at time $t_{2}$, followed by a second transfer $\pi$-pulse which swaps the two coherences between $\ket{s}$ and $\ket{d}$. The two coherences subsequently evolve such that they reach the same phase simultaneously. At this point, a transfer $\pi/2$-pulse is applied, implementing an effective $50{:}50$ beamsplitter between the two atomic coherences.

We note that the coherences need not be fully rephased when the $\pi/2$-pulse is applied; rather, it is sufficient that they have identical phase evolution for each velocity class. Following this beamsplitter operation, a transfer $\pi$-pulse is applied, causing one coherence to rephase at time $t_{3}$, where a readout control pulse retrieves the corresponding optical mode. A final transfer $\pi$-pulse rephases the second coherence, which is subsequently retrieved by a second control pulsea time $t_{4}$.

This sequence is equivalent to a Mach-Zehnder interferometer in the time domain. Such protocols could be used to perform measurements of time-bin qubits in different bases, highlighting the potential of the rephased ORCA memory as a platform not only for multimode storage, but also for high-bandwidth temporal-mode processing.

\end{document}